\documentclass{aa}
\usepackage{epsfig}
\usepackage{graphicx}
\begin{document}

\title{A boxy bulge in the Milky Way.\\ Inversion of the stellar
statistics equation with 2MASS data}

\author{M. L\'opez-Corredoira\inst{1,2}, A. Cabrera-Lavers\inst{1},
O. E. Gerhard\inst{2}}
\institute{
$^1$ Instituto de Astrof\'\i sica de Canarias, C/.V\'\i a L\'actea, s/n,
E-38200 La Laguna (S/C de Tenerife), Spain\\
$^2$ Astronomisches Institut der Universit\"at Basel,
Venusstrasse 7, CH-4102 Binningen, Switzerland}

\offprints{martinlc@iac.es}

\date{Received xxxx / Accepted xxxx}

\abstract{
Inverting the stellar statistics equation from 2MASS star counts, 
we obtain the 3D density distribution of the Galactic
bulge as well as its luminosity function in the K-band. 
This results in a boxy bulge with axial ratios 1:0.5:0.4 and a major axis angle 
with respect to the Sun-galactic center of $20^\circ-35^\circ $.
}

\authorrunning{L\'opez-Corredoira et al.}

\titlerunning{Bulge/2MASS}

\maketitle

\section{Introduction}

Due to  extinction, near-infrared (NIR) surveys are the most appropriate 
observations to study the stellar population of the Galactic central 
components. However, we see them edge-on, so the task of constraining their 
parameters is not as easy as in other external galaxies, which can be 
observed face-on. In order to recover the 3D information,
L\'opez-Corredoira et al. (2000; hereafter L00)  developed
a method of inversion of the star counts based on Bayesian statistics, 
which permits the obtaining of both
the stellar density distribution and the luminosity function (LF) without
making any a priori assumptions about them.

The L00 inversion technique has been applied to the K-band 
TMGS star counts up to $m_K=9.0$ in some strips that crossed the
Galactic plane, mostly in positive longitudes (L00). The inversion was also
applied to mid-infrared MSX Galactic plane maps 
(L\'opez-Corredoira et al. 2001a), although in
this last case there were very few stars to derive the stellar density and
only the LF could be derived with some accuracy.
Now, we have  a deeper and  larger coverage (the whole sky)
survey: 2MASS, with 3 NIR filters (which permits  extinction correction) as well as
 some improvements on the disc models (necessary to
be subtracted from the total star counts and isolate the bulge counts), so we
have the chance to obtain more precise results from the bulge density
distribution and its LF. This is the goal of the present
paper: using the L00 inversion technique with the newest 
data and information.

\section{2MASS star counts and disc subtraction}
\label{.disc}

The 2MASS survey provides magnitudes of all stars in the NIR filters
H, K$_s$ in the range $7.5<m_H<15.1$, $7.0<m_K<14.3$ and nearly one magnitude
less in the galactic plane (Skrutskie et al. 1997). It also provides J magnitudes, 
but we will not use these here. We can build
maps of cumulative star counts within these ranges (Alard 2001). In order to
make an approximate correction of the extinction (Alard 2001), we calculate
\begin{equation}
m_e=m_K-1.77(H-K)
\label{magcorr}
,\end{equation}
and we produce maps of cumulative star counts with $m_e<m_{e,0}$ and $m_{e,0}$ 
between 7.0 and 10.75 in steps of 0.25 mag. The angular space resolution 
both in longitude and latitude is 0.5$^\circ $ in the range 
$-20^\circ <l<20^\circ $, $-12^\circ <b<12^\circ $.

In order to isolate the bulge star counts, we must first
subtract the disc star counts in the regions where both disc and bulge 
contribute along the line of sight. We cannot apply the inversion of L00 to
disc+bulge counts since both components 
have different LFs and, moreover, the
L00 method does not work for stars near the Sun as the relative error
of the kernel in the inversion becomes very large.

The disc counts are given in L\'opez-Corredoira 
et al. (2004), interpreted as a truncated and/or flared disc:

\begin{description}

\item[Disc model 1:] An exponential function of the galactocentric distance 
($R$) for $R>4$ kpc:
\begin{equation}
\rho =\rho _\odot e^{-\left(\frac{R-R_\odot}{1970 \ {\rm pc}}\right)}
e^{-\frac{|z|}{h_z(R)}}\ {\rm star\ pc}^{-3}
,\end{equation}
\begin{equation}
h_z(R)=285e^{\left(\frac{R-R_\odot}{[12-0.6R({\rm kpc})] \ {\rm kpc}}\right)}{\rm pc}
,\end{equation}
and a constant density for $R\le 4$ kpc:
\begin{equation}
\rho =5.1\rho _\odot e^{-\frac{|z|}{h_z(R)}}\ {\rm star\ pc}^{-3}
,\end{equation}
\begin{equation}
h_z(R)=509-48R(\rm kpc) \ {\rm pc}
.\end{equation}

\item[Disc model 2:] A hole in the  inner disc which can be modeled $\forall R<R_\odot$ as:
\[
\rho =\left[\rho _\odot e^{\left(\frac{R_\odot}{1970\ {\rm pc}}
+\frac{3740\ {\rm pc}}{R_\odot}\right)}\right]e^{-\left(\frac{R}{1970\ {\rm pc}}
+\frac{3740\ {\rm pc}}{R}\right)}e^{-\frac{|z|}{h_z(R)}}
\]\begin{equation}
 {\rm star\ pc}^{-3}
,\end{equation}
\[
h_z(R)=285[1+0.21\ {\rm kpc}^{-1}(R-R_\odot)
\]\begin{equation}
+0.056\ {\rm kpc}^{-2}(R-R_\odot)^2] \ {\rm pc}
.\end{equation}

\end{description}
In all cases we adopt $R_\odot=7.9$ kpc. 
Indeed both models are almost equivalent for
2.5 kpc$<R<R_\odot $. They differ substantially only for $R<2.5$ kpc. 
We do not know the exact disc profile in this region where it is 
mixed with the bulge, although we know that it lies
between model 1 and 2. 

The disc LF used in the K-band is that given by Eaton et al. (1984). 
In order to obtain the correspondence
between $m_K$ and $m_e$, through eq. (\ref{magcorr}),
we calibrate $\langle (H-K)_0\rangle $ (color in regions without
extinction) in a region towards the center 
where the disc is almost isolated and without extinction
($l=20^\circ $, $b=6^\circ $), as a function of $m_{e,0}$. 
In the regions $-20.25^\circ<l<-17.75^\circ $, 
$3.75^\circ<|b|<12.25^\circ $, we calibrate the amplitude of 
the density which gives the best fit: $\rho _\odot=0.050$ star pc$^{-3}$ for 
both models. This is close to the value used in L\'opez-Corredoira et al. (2004). 

\begin{figure}
\begin{center}
\vspace{1cm}
\mbox{\epsfig{file=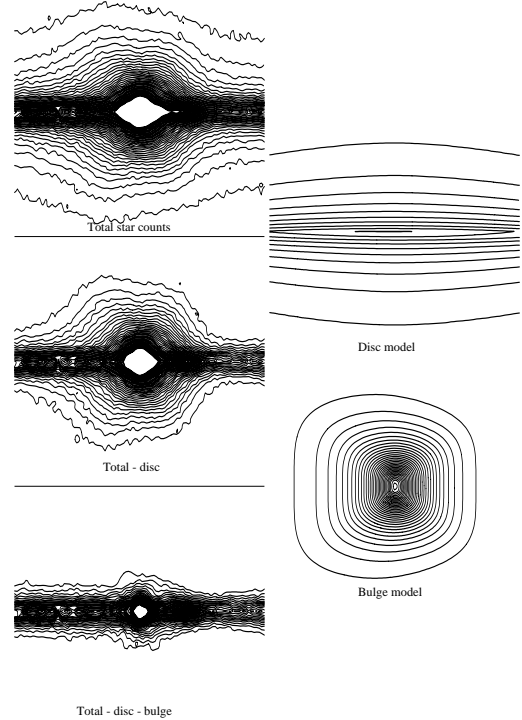,height=10cm}}
\end{center}
\caption{Star counts maps with $m_e<9.5$, for the region
$-20^\circ <l<20^\circ$, $-12^\circ<b<12^\circ$.
``Total'' map stands for the counts derived from 2MASS. 
``Disc'' map is the prediction of the disc ``model 1''
contribution (see \S \ref{.disc}). 
``Total-disc'' is the subtraction of the disc
in the total counts map. ``Bulge'' stands for the best triaxial
boxy-isodensity fit of the result of the inversion (see \S \ref{.boxy}) 
of ``Total-disc''. ``Total-disc-bulge'' are the residuals after subtracting 
the bulge from the ``Total-disc'' map. 
The isocounts contours represent, from the outermost to the innermost: 600 to 30000 star
deg$^{-2}$ with steps of 
600 star deg$^{-2}$.}
\label{Fig:maps}
\end{figure}

The results of the disc subtraction using model 1 are
shown in Fig. \ref{Fig:maps} (Total - disc) for $m_e<9.5$; for different
$m_{e,0}$ the results are similar.

\section{Inversion of bulge star counts}
\label{.3axbulge}

The fitting of parameters of the density and the LF
in the bulge would be a way to carry out the analysis, but many free
parameters would be needed, especially for the LF, and
a knowledge of the shape of both functions is required.
The L00 inversion technique is not limited in this way. 
However, some assumptions are needed:
the disc models in \S\ref{.disc} and the reddening law of eq. (\ref{magcorr}); 
the assumption of constant LF throughout the bulge, 
and that the noise effects in the inversion do not  significantly change the 
solution (as tested).

The procedure to invert the stellar statistics equation once the disc is
subtracted is  described in L00. It is an iteration of consecutive Lucy iterations.
The LFs in K-band and the density for each line of sight are obtained 
as numerical functions; nevertheless, 
we fit ``a posteriori'' the parameters of an analytical 3-D 
function for the density of a triaxial bulge, and we use 
a weighted average of LFs in the different lines of sight.
The result of each individual iteration does not depend significantly on
the initial guess and, in the final global outcome, the average/fit
of all the lines of sights, the dependence on the initial guesses is
even smaller, much smaller than other sources of error and dispersion with 
respect to the average/fit.  

There is a  wrong statement in L00. $R_\odot $ cannot be 
determined by this method, because any factor in the distance scale can
be compensated for by making the absolute magnitudes of all stars fainter
or brighter; we recently realized that the value $R_\odot =7.9$ kpc given
by L00 was indeed obtained because of the selected range of 
$M_K$ in the algorithm. This value was obtained by chance. However, the rest of the parameters are independent
of the assumed value of $R_\odot $; it only affects  the horizontal
calibration of the LF, and, since $R_\odot =7.9$ kpc is
in agreement with other measures, the presented results in L00 should be
correct. Thus, we continue to adopt $R_\odot =7.9$ kpc.

We repeat here  the steps explained by L00 but with the
present 2MASS data instead of TMGS data.
The selected region is $-20^\circ <l<20^\circ$, 
$2.25^\circ<|b|<12.25^\circ $; we avoid the Galactic plane, where other
possible sources apart from the disc and bulge may be present. 
The subtracted disc model is different from that of L00 (see \S \ref{.disc}).
A major improvement is that we do not need to specify  the
extinction, but we can correct the star counts by means of eq. (\ref{magcorr}).
By doing this, we avoid the possible problems with the patchiness of the
extinction. The selected range of magnitudes to obtain the 3D-density 
is: $7.00<m_{e,0}<9.50$. For the LF, 
the inversion is less sensitive to the errors, 
so we have  further extended the range up to $m_{e,0}<10.75$.
One cannot extend the fainter limit too much because
the ratio bulge/disc becomes poorer.
The normalization of the LF is made assuming
the value given by L00 for $\phi (M_K=-6.4)=2.34\times 10^{-5}$.
We get two outcomes: the LF in the range 
between $M_K=-6.6$ and $M_K=-2.8$, and the 3D-density
distribution of the bulge.

\begin{figure}
\begin{center}
\vspace{1cm}
\mbox{\epsfig{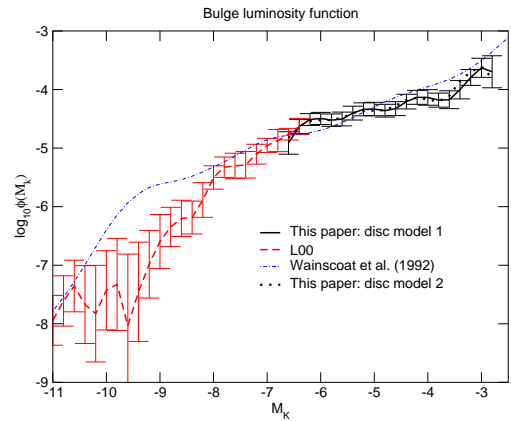}}
\end{center}
\caption{Luminosity function derived in this paper (with disc models
1 and 2; error bars correspond to model 1 and are similar for model 2) 
and L00 in comparison with the synthetic LF of Wainscoat 
et al. (1992). Assumed $R_\odot =7.9$ kpc.}
\label{Fig:lf}
\end{figure}

Fig. \ref{Fig:lf} shows the LF, with the extension
to brighter stars derived from the TMGS by L00. 
In total, we have a LF over 8 magnitudes. A comparison with Wainscoat et al.'s (1992)
synthetic LF (derived from a model of stellar populations)
shows that there is a strong deficit of very bright stars 
($M_K<-8$)---supergiants and late M giants---with respect to the synthetic 
LF (L00), and slight differences around $M_K=-6$ and
$M_K=-3.5$.

\begin{figure}
\begin{center}
\vspace{1cm}
\mbox{\epsfig{file=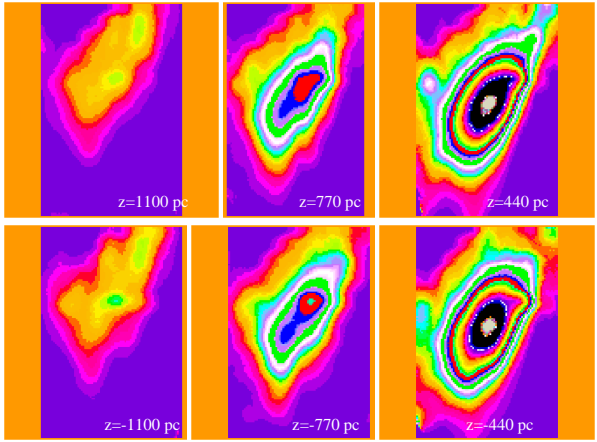,height=8cm,angle=-90}}
\end{center}
\caption{Different cuts ($z=\pm 440, \pm 770, \pm 1100$ pc) 
through the 3D bulge density distribution, obtained from the bulge counts with the
L00 inversion technique after subtracting the model 1 for the disc. 
Vertical axis, parallel
to the line Sun-Galactic center, between 
4.4 and 11.0 kpc from the Sun; increased distance from the Sun upwards. 
Horizontal axis, perpendicular to the
line Sun-Galactic center: between -2.2 and 2.2 kpc;
positive longitude to the left. $R_\odot =7.9$ kpc. The contours represent
isodensity regions; the density range between 0 and 1.4 star pc$^{-3}$.
Gaussian smoothing of the data was applied with $\sigma _x=\sigma _y=220$ pc,
$\sigma _z(\pm 440\ {\rm pc})=55$ pc, $\sigma _z(\pm 770\ {\rm pc})=110$ pc,
$\sigma _z(\pm 1100\ {\rm pc})=110$ pc.}
\label{Fig:inv}
\end{figure}

\begin{figure}[!h]
{\par\centering \resizebox*{2.8cm}{2.8cm}{\includegraphics{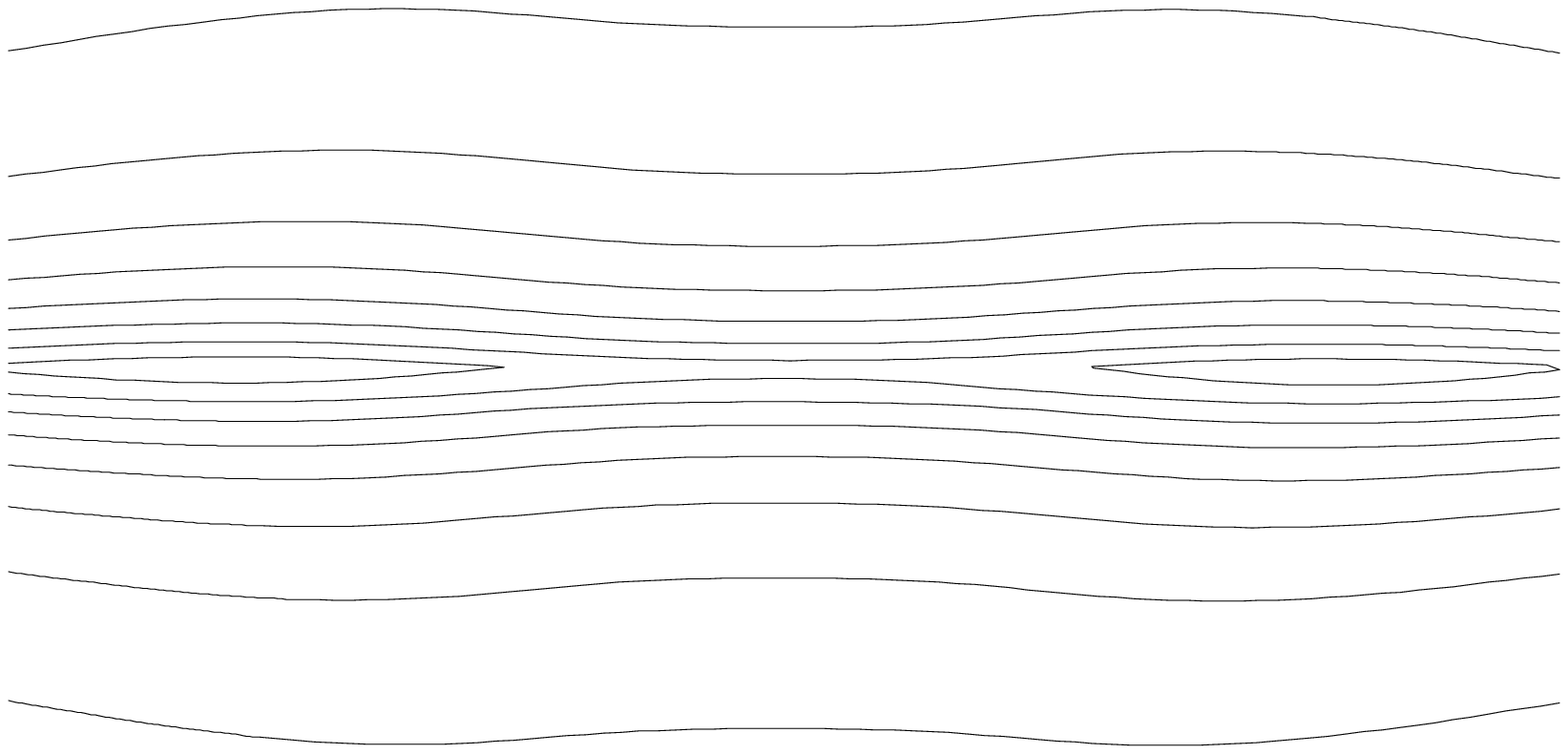}}
\resizebox*{2.8cm}{2.8cm}{\includegraphics{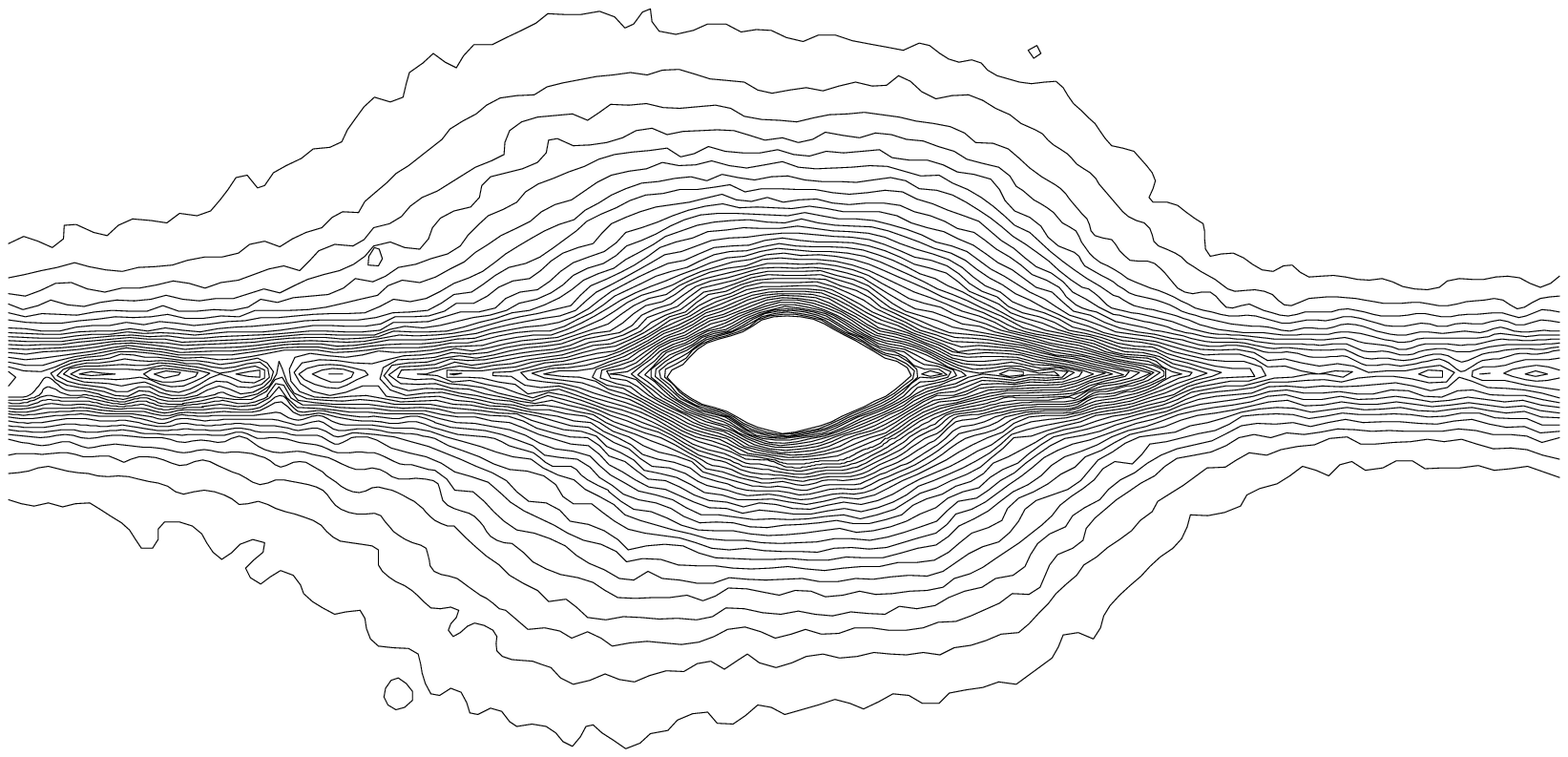}}
\resizebox*{2.8cm}{2.8cm}{\includegraphics{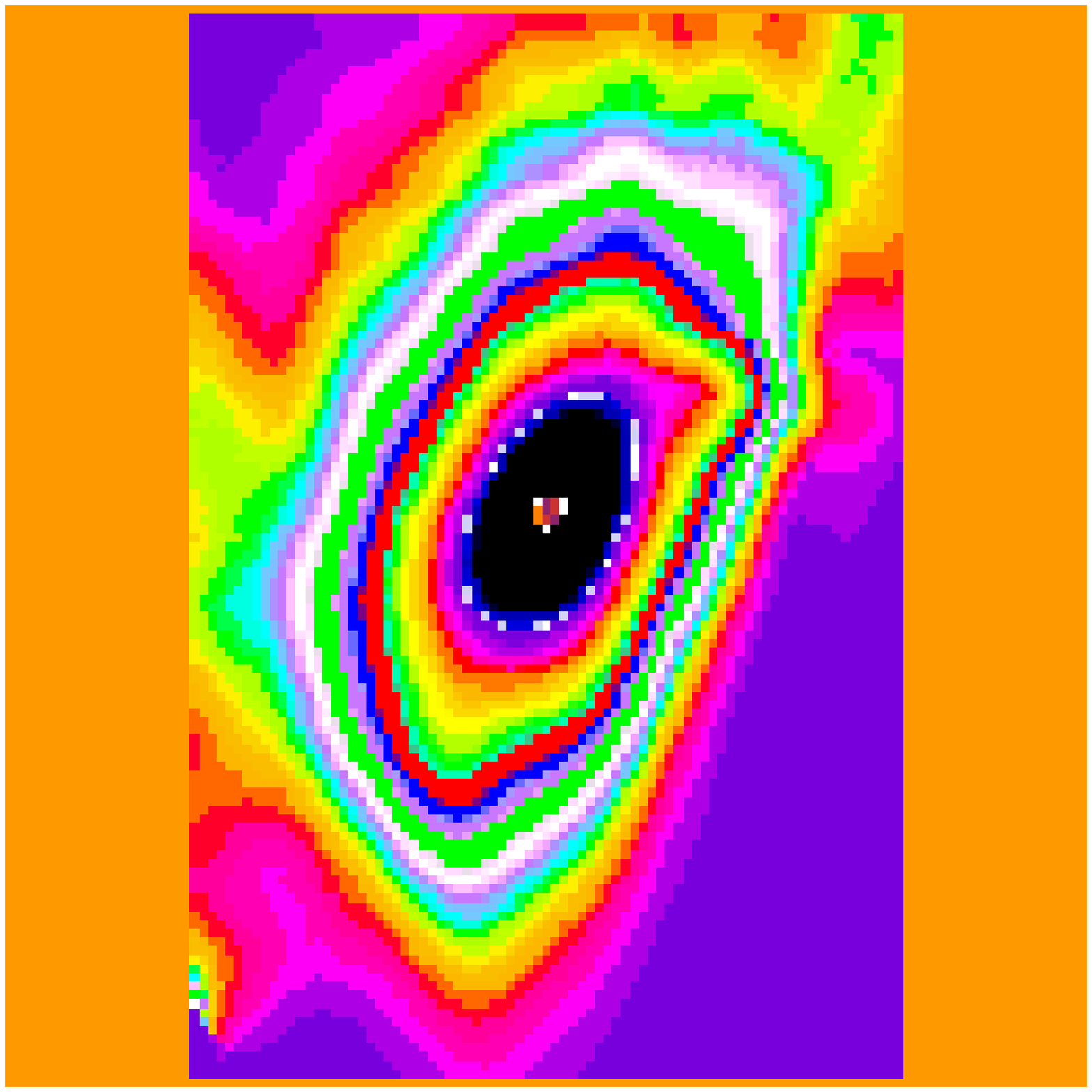}}\par}
\caption{Left: Star counts of the disc ``model 2''
contribution (see \S \ref{.disc}); 
isocounts contours like in Fig.
\protect{\ref{Fig:maps}}. Middle: ``Total-disc'' with ``model 2''.
Right: Cut at $z=-440$ pc 
as in Fig. \protect{\ref{Fig:inv}} but with ``model 2'' for the disc.}
\label{Fig:model2}
\end{figure}

Fig. \ref{Fig:inv} shows different cuts of the 3D density distribution
with $z=$constant, obtained when using model 1 of the disc. 
Model 2 of the disc gives similar results (see Fig. \ref{Fig:model2}).
This figure gives the smoothed raw result from the L00 inversion
technique, $\rho (\vec{r})$, and  clearly  shows
the non-circularity of the contours in the bulge cuts. 
This confirms that the bulge is non-axisymmetric, with
its elongated structure closer to us in the first quadrant 
($l>0$).

If we assume that the bulge is constituted by triaxial ellipsoids, their
parameters would be as follows: For the model 1 of the disc, the fitted mean 
axial ratios are 1:0.49:0.37; the angle between the major axis 
of the bulge (in the first quadrant) and the line 
Sun-Galactic center: $\alpha =29^\circ $; and the density profile 
$\rho(t)\approx 6.6e^{-\frac{t}{740\ {\rm pc}}}$ star pc$^{-3}$,
with $t=\sqrt{x_1^2+(x_2/0.49)^2+(x_3/0.37)^2}$
where $x_i$, $i=1,2,3$ are along the
major, middle and minor axis of the triaxial ellipsoid, respectively.
Tests to recover a given 3D-density/LF with known parameters 
applying the whole L00-method to its projected star counts show 
that typical systematic deviations due only 
to the inversion method and noise are: 
90 pc for the scale length of the profile, 8$^\circ $ for the angle, 
and 0.04 for both axial ratios. For the model 2 of the disc, 
the mean axial ratios are 1:0.49:0.40, angle $\alpha =27^\circ $.

The bulge is closer to axisymmetry (ratios 1:1:x) than 
1:0.33:0.22 (Dwek et al. 1995), 1:0.38:0.26 (Freudenreich 1998) or 
1:(0.3-0.4):0.3 (Bissantz \& Gerhard 2002).
However, it is not very far from the values of
1:0.6:0.4 (Binney et al. 1997), 1:0.43:0.29 
(Stanek et al. 1997) or 1:0.54:0.33 (L00). 
A lower ellipticity
is more in accord with  other galaxies, because  bulges generally do not
have large asymmetries. This also contributes to clarify
the difference between a bulge and a bar in our Galaxy. The present
structure looks like a bulge rather than a stick-like bar.
We suspect that the derived bulge axial ratios are 
different due to the dependence of the methods to correct for extinction
and the disc subtraction. In our case we used NIR star 
counts corrected for extinction based on reddening,
and an inner truncation and/or flare in the disc model increases 
the number of stars associated with the bulge, which reduces its 
global axisymmetry. The result might also depend on the
method of inversion/fitting and the a priori assumptions.
  
A parameter which seems to be in agreement with most other studies
is the orientation of the major axis of the bulge (when they refer to
the bulge and not to the hypothetical in-plane bar; see L\'opez-Corredoira
et al. 2001b): 12$^o$-30$^o$ (Dwek et al. 1995;
Nikolaev \& Weinberg 1997; Stanek et al. 1997; Binney et al. 1997; 
Freudenreich 1998; Sevenster et al. 1999; L00; Bissantz \& Gerhard 2002).

\section{Boxy Bulge}
\label{.boxy}

\begin{figure}[!h]
{\par\centering \resizebox*{2.8cm}{2.8cm}{\includegraphics{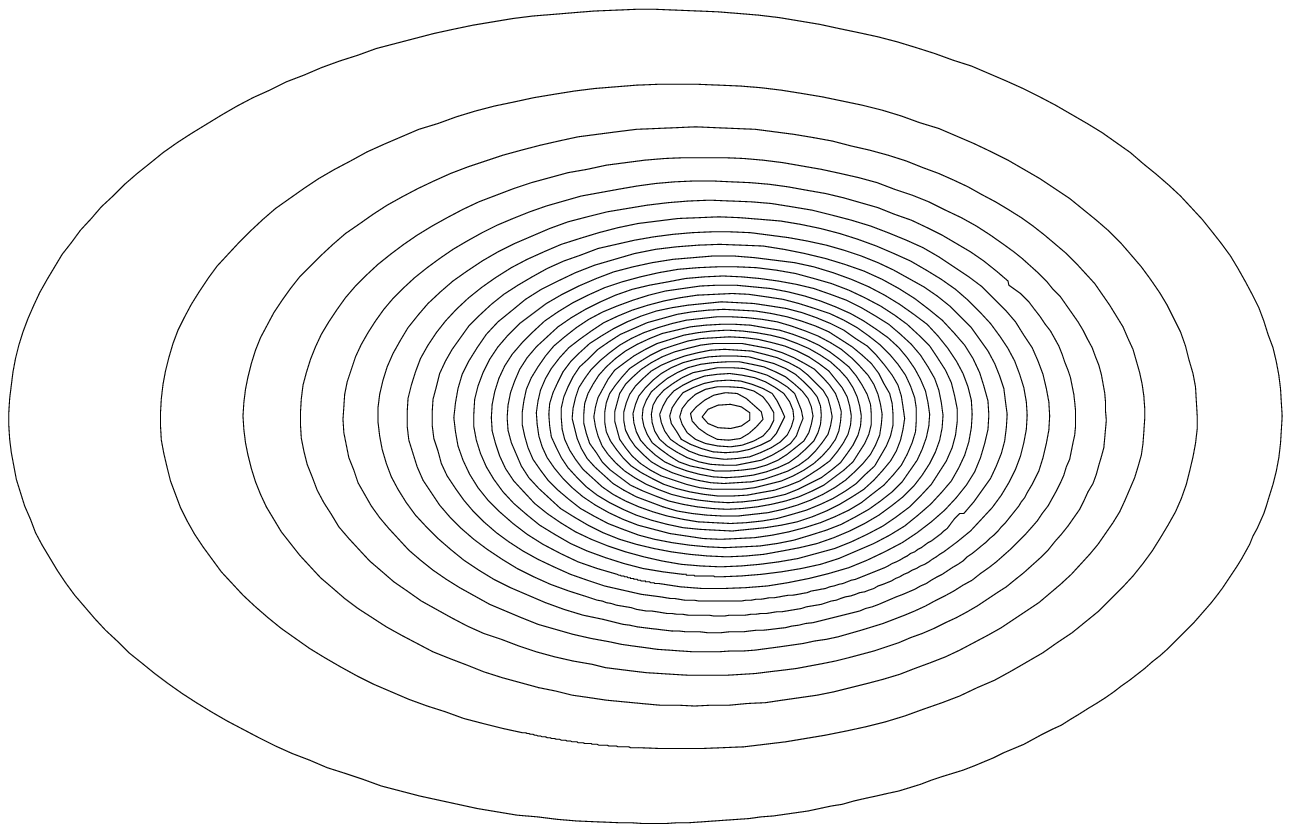}}
\resizebox*{2.8cm}{2.8cm}{\includegraphics{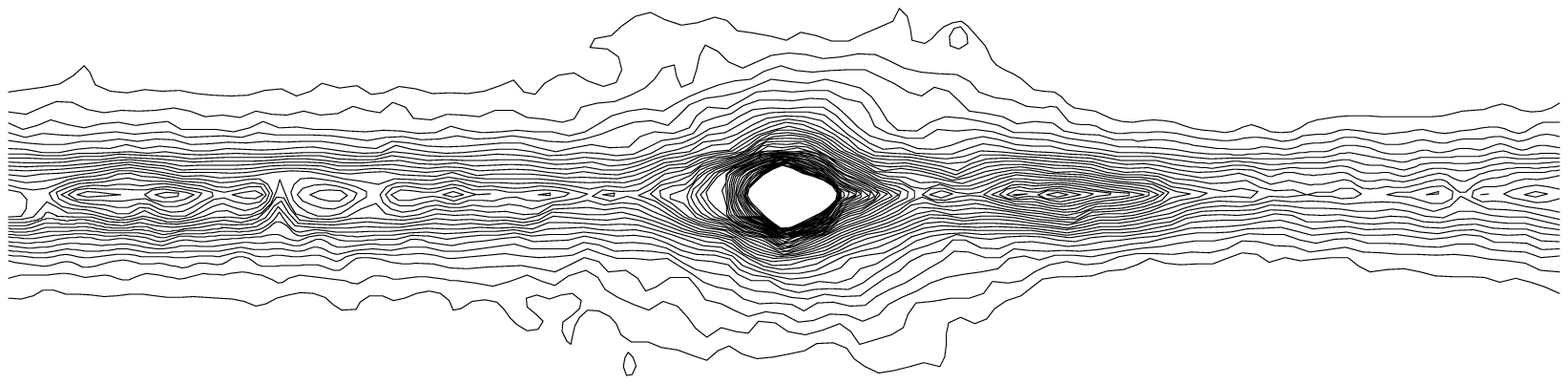}}\par}
\caption{Left: best triaxial
ellipsoid fit of the result of the inversion (see \S \ref{.3axbulge}) of 
``Total-disc'' in Fig. \protect{\ref{Fig:maps}}. 
Right: ``Total-disc-bulge'' map. Isocount contours as in Fig.
\protect{\ref{Fig:maps}}.}
\label{Fig:elip}
\end{figure}

Although a triaxial ellipsoid
represents the bulge in first order, the shape of the projection
(``Total-disc'' in Fig. \ref{Fig:maps}) is boxier than
the projection of ellipsoids (Fig. \ref{Fig:elip}).
The boxy shape of the Bulge was  noted previously
(Kent et al. 1991, 1992; Dwek et al. 1995). Here, the boxiness
is even more remarkable than in previous papers, due perhaps to
a different subtraction of the disc and the availability of
star counts instead of flux maps. The isodensity contours in
Fig. \ref{Fig:inv}, especially those at higher $t$ and smaller $z$,
look boxier also  than an ellipse.
If we use $t=[x^4+(x_2/a)^4+(x_3/b)^4]^{1/4}$---a similar boxy structure 
 as was proposed by Kent et al. (1991) for the 
axisymmetric case---we get a better representation
with somewhat fewer residuals (Fig. \ref{Fig:maps}). 
For this boxy-model, we adopt the same angle $\alpha $ and axial
ratios, and the same profile $\rho (t)$ as for the  ellipsoids
but with $t_{\rm boxy}=0.866t_{\rm ellip}$. This is the approximation
that gives the best correspondence between ellipsoids and boxy-isocontours
with the same inclination and axial ratios. The integrated
boxy-contours do not perfectly represent the observed isocount contours
in the outmost regions---it seems that the corners are not so 
abrupt---but they give a better model than the elliptical
bulge. The asymmetry of the counts in the 
residuals in plane regions (Fig. \ref{Fig:maps}, ``total-disc-bulge'')
suggests that part of these remaining stars are due to another
axisymmetric component, presumably a long bar. Other components like
a ring, spiral arms, and  errors in the disc subtraction and extinction
correction in plane regions are also expected to contribute to the residuals.

\section{Discussion and conclusions}

It is well known that boxy bulges are related to  bars 
in edge-on galaxies (Merrifield \& Kuijken 1999). 
Their bar-like nature is revealed by the spectroscopical data which show 
the typical dynamics of  multiple-component line emission with
one of the components being bar-like. Our Galaxy has
one of these boxy-bulges, which is not itself long-bar-like
but quite thick: axial ratios 1:0.5:0.4. Therefore, the scenario
of an in-plane bar apart from the bulge 
(L\'opez-Corredoira et al. 2001b, Hammersley et al.
2001, and references therein) is supported. 
The asymmetry of the residuals ``Total-disc-bulge'' also agrees with this scenario.
The triaxial bulge has an angle 
$\alpha \approx 20^\circ - 35^\circ $  with respect to the line Galactic center-Sun,
smaller than the long-bar angle.

\

Acknowledgments:
Thanks to the anonymous referee and Peter Englmaier for 
helpful comments. This publication makes use of data products from 
2MASS, which is a joint project of the Univ. of Massachusetts and the 
Infrared Processing and Analysis Center (IPAC), funded by the NASA and 
the NSF. We thank the Swiss National Science Foundation for support under 
grant 20-64856.01

\end{document}